\documentclass[letterpaper, 10 pt, conference]{ieeeconf}  

\IEEEoverridecommandlockouts                              

\usepackage[pdftex]{graphicx}
\usepackage{algorithm}
\usepackage{algpseudocode}

\usepackage{amssymb,amsmath, amsthm}
\usepackage{subfig}
\usepackage{indentfirst}
\usepackage{color}
\usepackage[hidelinks]{hyperref}
\usepackage{enumerate}
\usepackage{tikz}
\usetikzlibrary{shapes,arrows}
\usepackage{epstopdf}
\usepackage{epsfig}
\usepackage{wrapfig}
\usepackage[font=small]{caption}
\usepackage{balance}
\usepackage{setspace}

\def\mf{\mathbf}
\def\mb{\mathbb}
\def\mc{\mathcal}
\def\beq{\begin{equation*}}
\def\eeq{\end{equation*}}
\def\bql{\begin{equation}}
\def\eql{\end{equation}}
\def\bqn{\begin{eqnarray*}}
\def\eqn{\end{eqnarray*}}
\def\bnl{\begin{eqnarray}}
\def\enl{\end{eqnarray}}
\def\bna{\bql\begin{array}{rcl}}
\def\ena{\end{array}\eql}
\def\bnn{\beq\begin{array}{rcl}}
\def\enn{\end{array}\eeq}
\def\bma{\begin{bmatrix}}
\def\ema{\end{bmatrix}}
\def\bmx{\begin{matrix}}
\def\emx{\end{matrix}}
\def\ben{\begin{enumerate}}
\def\een{\end{enumerate}}
\def\bit{\begin{itemize}}
\def\eit{\end{itemize}}
\def\bei{\begin{itemize}}
\def\eei{\end{itemize}}
\def\bet{\begin{tabular}}
\def\eet{\end{tabular}}
\def\span{\text{span}}

\newcommand{\allcaps}[1]{\uppercase\expandafter{#1}}

\providecommand{\norm}[1]{\left\|#1\right\|}

\theoremstyle{definition}

\newtheorem{remark}{Remark}

\makeatletter
\let\OldStatex\Statex
\renewcommand{\Statex}[1][3]{%
  \setlength\@tempdima{\algorithmicindent}%
  \OldStatex\hskip\dimexpr#1\@tempdima\relax}
\makeatother

\allowdisplaybreaks
\title{\LARGE \bf
Sequential Learning from Noisy Data: Data-Assimilation Meets Echo-State Network
}

\author{Debdipta Goswami
\thanks{Debdipta Goswami is an Assistant Professor in the Department of Mechanical and Aerospace Engineering, The Ohio State University, Columbus, OH 43210.
        {\tt\small goswami.78@osu.edu}}%
}

\begin{document}
\maketitle

\tikzstyle{block} = [draw, fill=blue!20, rectangle, 
    minimum height=3em, minimum width=4em]
\tikzstyle{sum} = [draw, fill=blue!20, circle, node distance=1cm]
\tikzstyle{input} = [coordinate]
\tikzstyle{noise} = [coordinate]
\tikzstyle{output} = [coordinate]
\tikzstyle{pinstyle} = [pin edge={to-,thin,black}]

\thispagestyle{empty}
\pagestyle{empty}

\begin{abstract}
This paper explores the problem of training a recurrent neural network from noisy data. While neural network based dynamic predictors perform well with noise-free training data, prediction with noisy inputs during training phase poses a significant challenge. Here a sequential training algorithm is developed for an echo-state network (ESN) by incorporating noisy observations using an ensemble Kalman filter. The resultant Kalman-trained echo-state network (KalT-ESN) outperforms the traditionally trained ESN with least square algorithm while still being computationally cheap. The proposed method is demonstrated on noisy observations from three systems: two synthetic datasets from chaotic dynamical systems and a set of real-time traffic data.

\end{abstract}

\section{INTRODUCTION}


The current interest in data-driven modeling and forecast of complex systems has motivated a wide array of machine-learning algorithms for different problems, e.g., classification, speech recognition \cite{Hinton2012}, board games \cite{Silver2016}, and even discovering mathematical algorithms \cite{Fawzi2022}. In particular, recurrent neural networks (RNNs) have proved to be useful in dynamical systems and time-series prediction. For example, an echo-state network (ESN) \cite{Jaeger2004} can model chaotic systems very effectively \cite{Lu2017}, \cite{Pathak2018}. However, these prediction techniques rely on relatively noise free training data in order to effectively tune the model parameters. But in many practical cases, the training data is noisy. Such scenarios include atmospheric modeling and traffic systems.

Neural network predictors, instead of using a physics-based handcrafted dynamic model, utilize the rich training dataset to build a parametric surrogate model, and then utilizes it to predict the future states. An ESN is a special type of RNN that uses a reservoir of nonlinear, randomly connected neurons to process time-varying input signal. Such a network with a convergence property, known to the ESN literature as echo-state property (ESP), can uniformly approximate any nonlinear fading memory filter \cite{Ortega2018}. The ESN is attractive as a neural predictor due to its ability to be tuned via output connections (also called the readout map) with minimal computing resources. Also,  a reservoir can be directly implemented by hardwares using field programmable gate arrays (FPGAs) or a photonic reservoir, thereby increasing efficiency and reducing computational overhead \cite{Tanaka2019}, \cite{Nakajima} . It is also extended to quantum computing realm via quantum reservoir computers (QRCs) \cite{Fuji2017}, \cite{Chen2019}. The effectiveness of ESN-based approaches for sparse estimation of chaotic systems and traffic network prediction is shown in \cite{Lu2017}, \cite{Goswami2021}, and \cite{Goswami2022}. However, the training of an ESN implicitly assumes a high fidelity training dataset representing the actual system states and therefore sensitive to the measurement noise. Although ESN-based methods \cite{Goswami2021} are developed to utilize noisy measurements to predict the unmeasured states during the testing phase, they rely on relatively noise free data for training.

On the other hand, in a model-based estimation problems from noisy measurements, the state estimate is computed in two steps. First, the motion update step facilitated by the model yields a forecast estimate. Then, the final estimate is generated by assimilating a noisy measurement via Bayesian update. For a linear system with Gaussian process and measurement noise, the optimal estimator is given by the celebrated Kalman filter \cite{Kalman}, whereas for a nonlinear system, optimal filtering is usually infinite dimensional requiring the solution of a stochastic partial differential equation. A variety of suboptimal finite dimensional techniques are usually employed for nonlinear estimation, e.g., the extended Kalman filter (EKF) \cite{Khalil}, unscented Kalman filter (UKF) \cite{Julier2004}, ensemble Kalman filter (EnKF),  and particle filter. However, these methods require a dynamic model to perform the motion update of the state estimate. 

This paper develops a sequential training method for an ESN by combining its strength of modeling an unknown dynamical system with a nonlinear filtering algorithm. An ESN is particularly well suited for modeling a dynamical system with faster and cheaper training that does not require backpropagation through time (BPTT). An ESN with fading memory can universally model nonlinear dynamics \cite{Maass2004}, \cite{Ortega2018}. The ESN architecture adopts an input-output neural network with a randomly generated recurrent reservoir where only output layer is trained.  The training is usually done via a least square linear regression that implicitly assumes noise free training data. The proposed method transforms the training problem of an ESN to a combined state-parameter estimation problem where the output weights are sequentially updated with the incoming noisy observations along with the system states. 

The combined estimation problem is, however, no longer linear. Hence, a nonlinear data-assimilation method is required for the measurement update. While the EKF and UKF perform well in model-based scenarios, the computation of the linearized dynamics is challenging for an ESN. The ensemble Kalman filter \cite{Bengtsson2003} is thus chosen for the measurement update for its strength in representing the posterior distribution of states by its sample means and covariances. The resulting algorithm is called Kalman training of the echo-state network (KalT-ESN). The algorithm is also extended to partial noisy measurements as training data by using a delay-embedding at the input layer \cite{Goswami2023}.   


The contributions of this paper are (1) providing a sequential training algorithm for an ESN via combined state and parameter estimation when the training data is noisy; (2) combining the prediction power of a recurrent neural network with the traditional Bayesian measurement update model of an ensemble Kalman filter; (3) improving the prediction accuracy over time for a chaotic nonlinear system from its noisy measurements; (4) extension of the algorithm for noisy partial measurements as the training data by delay-embedding; and (5) application of the proposed method to a real set of mobility data in order to predict daily cycles of traffic volume.  The model-free estimation algorithm developed here has wide applications for estimation of complex dynamics from noisy observations when a reliable model is unavailable.

The paper is organized as follows. Section II provides a brief overview of the echo-state network (ESN). Section III presents the combined state and parameter estimation problem and the KalT-ESN algorithm. Section IV illustrates the applications to three different problems: two synthetic data streams generated by chaotic nonlinear systems and one real set of data traffic sensor data. Section V concludes the manuscript and discusses ongoing and future work.

\section{Echo-State Networks: A Universal Predictor}
Echo-state networks are a type of recurrent neural network used for dynamical systems prediction. It consists of a large dynamic reservoir of randomly connected neurons driven nonlinearly by input signals. These neuronal responses are then linearly combined to match a desired output signal. Due to its dependence on the richness of the dynamic reservoir, it is also called a reservoir computer (RC). An ESN consists of an input layer $\mf{u}\in\mb{R}^m$, coupled through input coupling matrix $W_{in}\in \mb{R}^{n\times m}$ with a recurrent nonlinear reservoir $\mf{r} \in \mathbb{R}^n$. The output $\mathbf{y}\in \mathbb{R}^p$ is generated from $n$ neurons of the reservoir via a readout matrix $W_{out}\in \mb{R}^{n\times p}$. The reservoir network evolves nonlinearly in following  fashion \cite{Maass2004}, \cite{Goswami2021}
\bql
\mf{r}(t+\Delta t)=(1-\alpha)\mf{r}(t) + \alpha\psi(W\mf{r}(t)+W_{in}\mf{u}(t)).
\eql
The time-step $\Delta t$ denotes the sampling interval of the training data. The leakage rate parameter $\alpha \in (0,1]$ helps slowing down the evolution of the reservoir states as $\alpha\rightarrow 0$. The nonlinear activation function $\psi(\cdot)$ is usually a sigmoid function, e.g., $\tanh(\cdot)$. The output $\mf{y}(t)$ is linearly read out from the reservoir states \cite{Maass2004}, \cite{Goswami2021}, i.e.,
\bql
\mf{y}(t)=W_{out}\mf{r}(t).
\eql
The weights $W_{in}$ and $W$ are initially randomly drawn and then held fixed. The weight $W_{out}$ is adjusted during the training process. The reservoir weight matrix $W$ is usually kept sparse for computational efficiency. 

During the training phase, an ESN is driven by an input sequence $\{\mf{u}(t_1),\ldots,\mf{u}(t_N)\}$ that yields a sequence of reservoir states $\{\mf{r}(t_1),\ldots,\mf{r}(t_N)\}$. The reservoir states are stored in a matrix $\mf{R}=[\mf{r}(t_1),\ldots,\mf{r}(t_N)]$. The correct outputs $\{\mf{y}(t_1),\ldots,\mf{y}(t_N)\}$, which are part of the training data, are also arranged in a matrix $\mf{Y}=[\mf{y}(t_1),\ldots,\mf{y}(t_N)]$.  The training is carried out by a linear regression with Tikhonov regularization as follows \cite{Jaeger2004}:
\bql \label{Eq: LSTrain}
W_{out} = \mf{Y}\mf{R}^T(\mf{R}\mf{R}^T + \beta\mf{I})^{-1},
\eql
where $\beta>0$ is a regularization parameter that ensures non-singularity. 
\begin{figure*}[t]
\centering 
\subfloat[]{\includegraphics[trim=0cm 0cm 0cm 0cm, clip=true, width=0.31\textwidth]{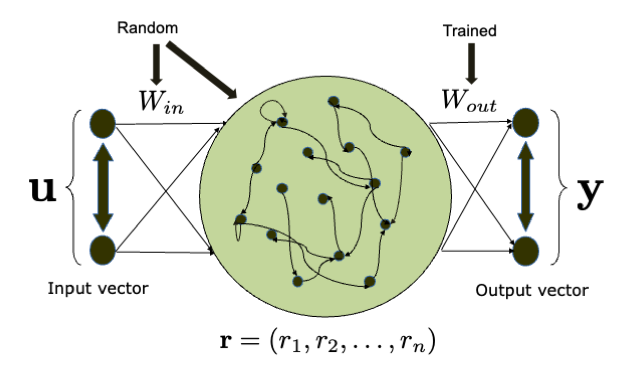}}
\subfloat[]{\includegraphics[trim=0cm 0cm 0cm 0cm, clip=true, width=0.35\textwidth]{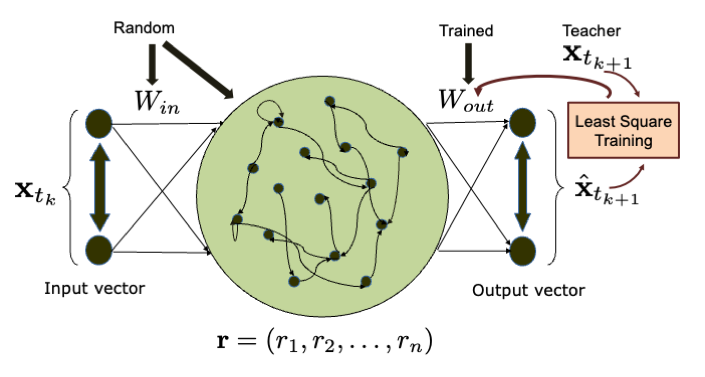}}
\subfloat[]{\includegraphics[trim=0cm 0cm 0cm 0cm, clip=true, width=0.34\textwidth]{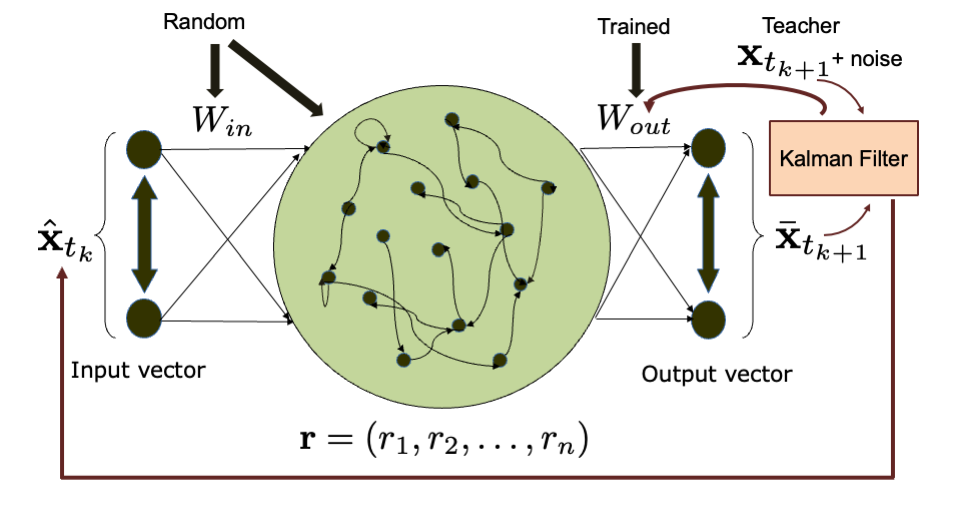}}
\caption{Architecture and training of an ESN: (a) the basic ESN, (b) least square training from a batch of noiseless data, and (c) Kalman filter based sequential training from noisy data } \label{Fig: Reservoir}
\end{figure*}

\begin{remark}
For an ESN to be a universal approximator, i.e., to be able to realize any nonlinear operator with bounded memory arbitrarily accurately, it must satisfy the echo-state property (ESP) \cite{Jaeger2004}. An ESN is said to have the ESP if the reservoir asymptotically washes out any information from the initial conditions. For the $\tanh(\cdot)$ activation function, it is empirically observed that the ESP holds for any input if the spectral radius of $W$ is smaller than unity \cite{Jaeger2004}. To ensure this condition, $W$ is normalized by its spectral radius. 
\end{remark}

\section{Sequential Training of an Echo-State Network: An Ensemble Kalman Filter Approach}
An ESN can be trained to predict a time-series $\{\mf{x}(t_i)\in \mb{R}^d:i\in\mb{N}\}$ generated by a dynamical system by setting $\mf{u}(t)$ and $\mf{y}(t)$ as the current and next state value (i.e., $\mf{x}(t_k)$ and $\mf{x}(t_{k+1})$) respectively. The network is trained for a certain training length $N$ of the time-series data $\{\mf{x}(t_i,i=1,\ldots,N\}$, which can then run freely by feeding the output $\mf{y}(t_k)$ back to the input $\mf{u}(t_{k+1})$ of the reservoir. In this case, both $\mf{u}$ and $\mf{y}$ have the same dimension $d$ as that of the time-series data. This setup is shown in Fig.~\ref{Fig: Reservoir}(a)-(b).

\begin{algorithm}
\caption{KalT-ESN: Sequential Training of an ESN with EnKF}\label{Alg: KalTESN}
\textbf{Input:} Noisy training data $\{\mf{y}(t_1),\ldots,\mf{y}(t_N)\}$, $\mf{y}(t_i) \in \mb{R}^d$\\
\textbf{Hyperparameters:} Training length $N$, leaking rate $\alpha$, regularization parameter $\beta$, reservoir connection probability $p\in(0,1)$, reservoir size $n$, activation $\psi$, ensemble size $M$, initial data covariance $\Sigma_{x}\in\mb{R}^{d \times d}$, initial weight covariance $\Sigma_w \in \mb{R}^{dn \times dn}$, measurement noise covariance $\Sigma_v \in \mb{R}^{d \times d}$\\
\textbf{Output:} $W_{in}$, $W$, $W_{out}$
\begin{algorithmic}[1]
\Procedure {Train}{ $\{\mf{y}(t_1),\ldots,\mf{y}(t_N)\}$; $\alpha$, $\beta$, $p$, $n$, $\psi$, $M$, $\Sigma_{x}$, $\Sigma_w$, $\Sigma_v$}
\State Generate $W\sim G(n,p)$ \Comment{Adjacency matrix of an Erd\"os-Renyi random graph}
\State Generate $W_{in}\in \mb{R}^{n\times d}$ random matrix
\State Generate $\hat{\mf{X}}(t_1) = [\hat{\mf{x}}^{(1)}(t_{1}),\ldots,\hat{\mf{x}}^{(M)}(t_{1})]$ \Statex $\sim \mc{N}(\mf{y}(t_1), \Sigma_x)$ \Comment{Data ensemble generation}
\State 	Generate $\mf{R}(t_1) = [\mf{r}^{(1)}(t_1), \ldots \mf{r}^{(M)}(t_1)] = \mf{0}_{n\times M}$
\State Generate $\hat{\mf{W}}_{o}(t_1) = [\underline{\hat{\mf{w}}}^{(1)}_{o}(t_1),\ldots,\underline{\hat{\mf{w}}}^{(M)}_{o}(t_1)]$ \Statex $ \sim \mc{N}(\mf{0}_{dn}, \Sigma_w)$ \Comment{Weight ensemble generation}
\For{$k=1$ to $N-1$}
\For {$i=1$ to $M$}
\State $\hat{W}_o^{(i)}(t_k) \gets \operatorname{reshape}(\underline{\hat{\mf{w}}}^{(i)}_{o}(t_k), d, n)$ 
\State $\overline{\mf{x}}^{(i)}(t_{k+1}) \gets \hat{W}_{o}^{(i)}(t_k)\psi(W\mf{r}^{(i)}(t_k) $ \Statex $ + W_{in}\hat{\mf{x}}^{(i)}(t_k))$ 
\EndFor
\State $\mf{X}^f(t_{k+1}) \gets [\overline{\mf{x}}^{(1)}(t_{k+1}),\ldots,\overline{\mf{x}}^{(M)}(t_{k+1})]$
\State $\mf{W}^f_o(t_{k+1}) \gets \mf{W}_o(t_k)$
\State $\mf{E}_w(t_{k+1}) = (\mf{W}^f_o(t_{k+1})-\bar{\mf{W}}^f_o(t_{k+1}))$ 
\State $\mf{E}_x(t_{k+1}) = (\mf{X}^f(t_{k+1})-\bar{\mf{X}}^f(t_{k+1}))$
\State $P_{wx}(t_{k+1}) \gets \mf{E}_w(t_{k+1})\mf{E}_w(t_{k+1})^T/(M-1)$
\State $P_{xx}(t_{k+1}) \gets \mf{E}_x(t_{k+1})\mf{E}_x(t_{k+1})^T/(M-1)$
\State $K^x(t_{k+1}) \gets P_{xx}(t_{k+1})\left[P_{xx}(t_{k+1}) + \Sigma_v\right]^{-1}$ \label{state_update}
\State $K^w(t_{k+1}) \gets P_{wx}(t_{k+1})\left[P_{xx}(t_{k+1}) + \Sigma_v\right]^{-1}$ \label{weight_update}
\State $\hat{\mf{X}}(t_{k+1}) \gets \mf{X}^f(t_{k+1})$ \Statex $ + K^x(t_{k+1})\left(\mf{y}(t_{k+1})-\mf{X}^f(t_{k+1})\right)$
\State $\hat{\mf{W}}_{o}(t_{k+1}) \gets \mf{W}^f_{o}(t_{k+1})$ \Statex $ + K^w(t_{k+1})\left(\mf{y}(t_{k+1})-\mf{X}^f(t_{k+1})\right)$
\EndFor
\State $W_{out} \gets  \operatorname{reshape}\left(\bar{\hat{\mf{W}}}_o(t_N), d, n\right)$\Comment{Final output weights}
\EndProcedure
\end{algorithmic}
\end{algorithm}`

An ESN proves to be a powerful tool for dynamical systems prediction when trained with noiseless data \cite{Lu2017}, \cite{Pathak2018}. Its performance is significantly improved when partial observations are available during the testing phase by assimilating them through an ensemble Kalman filter \cite{Goswami2021}. It can also be modified to accommodate partial state measurements as training data by a higher dimensional delay-embedding in the input layer \cite{Goswami2023}. However, a significant challenge is posed when noisy data is present during the training phase. A simple approach is to ignore the noise and train the network using the least square method \eqref{Eq: LSTrain}.  But this simplistic approach disregards the fact that the training data might contain rapid variations and outliers that might impart large errors in the weights of the ESN. The problem becomes particularly worse for chaotic time-series where the training noise can drive the network away from the chaotic attractor. 

This paper proposes an alternative method for training the ESN from noisy data by incorporating a filtering step recursively in the training process. We assume that the training data $[\mf{y}(t_1),\ldots, \mf{y}(t_N)]$ can be modeled as $\mf{y}(t_k) = \mf{x}(t_k) + \mf{v}_k$ with $\mf{v}_k$ as a zero-mean i.i.d. Gaussian random variable with covariance $\Sigma_v$. A combined state and parameter estimation method is then used via state augmentation. The forecast model is formulated as
\bnl \label{Eq: Forecast}
\mf{r}(t_{k+1}) &=& (1-\alpha)\mf{r}(t_k) \\\nonumber &&+ \alpha\psi\left(W\mf{r}(t_k)+W_{in}\hat{\mf{x}}(t_k)\right)\\\nonumber
\overline{\mf{x}}(t_{k+1}) &=& \hat{W}_{out}(t_k)\mf{r}(t_{k+1}) \\\nonumber
\overline{W}_{out}(t_{k+1}) &=& \hat{W}_{out}(t_k)
\enl
where $\mf{r}(t_k)$ is the known reservoir state, $\hat{\mf{x}}(t_k)$ and $\hat{W}_{out}(t_k)$ are the estimates of the true data and output weight at time $t_k$. Now, since we can observe the true $\mf{r}(t_k)$, we can define $\Phi^x_{t_k}(\cdot) \triangleq (1-\alpha)\mf{r}(t_k) + \alpha\psi(W\mf{r}(t_k)+W_{in}(\cdot))$ so that the forecast model \eqref{Eq: Forecast} becomes

\bnl \label{Eq: ForecastModified}
\overline{\mf{x}}(t_{k+1}) &=& \hat{W}_{out}(t_k)\Phi^x_{t_k}(\hat{\mf{x}}(t_k)) \\\nonumber
\overline{W}_{out}(t_{k+1}) &=& \hat{W}_{out}(t_k),
\enl
with $\overline{\mf{x}}(t_{k+1})$ and $\overline{W}_{out}(t_{k+1})$ as the data and output weight forecast at $t_{k+1}$ respectively. Thus, we now have a dynamic forecast model \eqref{Eq: ForecastModified} which can be formulated into a Kalman filter step for the analysis part. First, we vectorize the $W_{out}$ matrix as $\underline{\mf{w}}_o \in \mb{R}^{nd}$ and define the new augmented state variable $\mf{z} = \begin{bmatrix}\mf{x}^T & \underline{\mf{w}}_o^T\end{bmatrix}^T$. The forecast model \eqref{Eq: ForecastModified} with this augmented variable $\mf{z}$ becomes  
\bnl \label{Eq: Forecastz}
\overline{\mf{z}}(t_{k+1}) &=& \Phi_{t_k}(\hat{\mf{z}}(t_k)),
\enl
where\[\Phi_{t_k}(\mf{z}) = \begin{bmatrix}
W_{out}\Phi^x_{t_k}({\mf{x}}) \\ \underline{\mf{w}_o}
\end{bmatrix},\] with $W_{out}$ as $\underline{\mf{w}}_o$ reshaped into a $d\times n$ matrix.
Since this is a nonlinear forecast model, a nonlinear modification of the Kalman filter algorithm is employed to jointly estimate the true training data $\mf{x}(t_k)$ and the output weight $W_{out}$. In particular, an ensemble Kalman filter (EnKF) is utilized to estimate the covariance of $\mf{z}(t_k)$ from a Gaussian ensemble realization. An initial ensemble $\hat{\mf{Z}}(t_1) = [\hat{\mf{z}}^{(1)}(t_1),\ldots \hat{\mf{z}}^{(M)}(t_1)]$ is generated with members drawn from $\mc{N}\left([\mf{y}(t_1)^T\,\,\, \mf{0}_{nd}^T]^T, \Sigma_z\right)$. The initial covariance $\Sigma_z$ is a hyperparameter and assumed to be block diagonal, i.e., $\Sigma_z = \operatorname{bdiag}(\Sigma_x, \Sigma_w)$. Algorithm \ref{Alg: KalTESN} presents the procedure for the sequential training of an ESN with EnKF. The inputs are the noisy training data $[\mf{y}(t_1),\ldots,\mf{y}(t_N)]$. After $N$ training steps, the procedure returns the trained ouput matrix $W_{out}$.

\begin{remark}
A block diagonal $\Sigma_z$ with only direct measurements of the state $\mf{x}$ explicitly separates the Kalman update step between the state $\mf{x}$ and parameters $\underline{\mf{w}}_o$ as shown in Algorithm \ref{Alg: KalTESN}, line \ref{state_update}-\ref{weight_update}.
\end{remark}

\begin{remark}
In addition to noisy training data, Algorithm \ref{Alg: KalTESN} works with sequentially available data in a recursive fashion, unlike the static least square training \eqref{Eq: LSTrain}.
\end{remark}

\begin{remark}
The Kalman based training of an ESN can assimilate the partial state measurements during the testing phase to keep improving the output weights $W_{out}$ and the subsequent time series estimates $\hat{\mf{x}}(t_k)$.
\end{remark}

\section{Numerical Examples}
This section illustrates the performance and ablation study of a the KalT-ESN algorithm on three time series data corrupted with noise. The first two are time-series generated by chaotic dynamical systems and the last one is a real-time traffic flow data obtained by Numina sensor nodes \cite{Numina} installed on the University of Maryland campus.
\begin{table*}[t]
\begin{center}
\caption{ESN hyperparameters}\label{tb:hyparam}
\begin{tabular}{lccc}
\hline
Hyperparameter & &Value &  \\
 & Lorenz system \eqref{Eq: Lorenz} & R\"{o}ssler system \eqref{Eq: Rossler} & Traffic Volume \\\hline
 Time step $\Delta t$ & $0.02$s & $0.1$s & $1$h\\
Reservoir size $n$ & $500$ & $500$ & $4000$\\
Reservoir connection probability $p$ & $0.01$ & $0.01$ & $0.01$ \\
Training length $N$ & $6000$ & $1000$ & $500$ \\ 
Activation $\psi(\cdot)$ & $\tanh(\cdot)$ & $\tanh(\cdot)$ & $\tanh(\cdot)$\\
Leaking rate $\alpha$ & $0.3$ & $0.3$ & $0.7$\\
Regularization $\beta$ & $10^{-6}$ & $10^{-6}$ & $10^{-6}$\\
Ensemble size $M$ & $300$ & $300$ & $300$\\
Initial state covariance $\Sigma_x = \sigma_x^2 I$ & $\sigma_x^2 = 0.2$ & $\sigma_x^2 = 0.2$ & $\sigma_x^2 = 10$\\
Initial weight covariance $\Sigma_w = \sigma_w^2 I$ & $\sigma_w^2 = 0.2$ & $\sigma_w^2 = 0.2$ & $\sigma_w^2 = 1$\\\hline
\end{tabular}
\end{center}
\end{table*}
\begin{figure}[t]
\centering 
\subfloat[]{\includegraphics[trim=0cm 0cm 0cm 0cm, clip=true, width=0.25\textwidth]{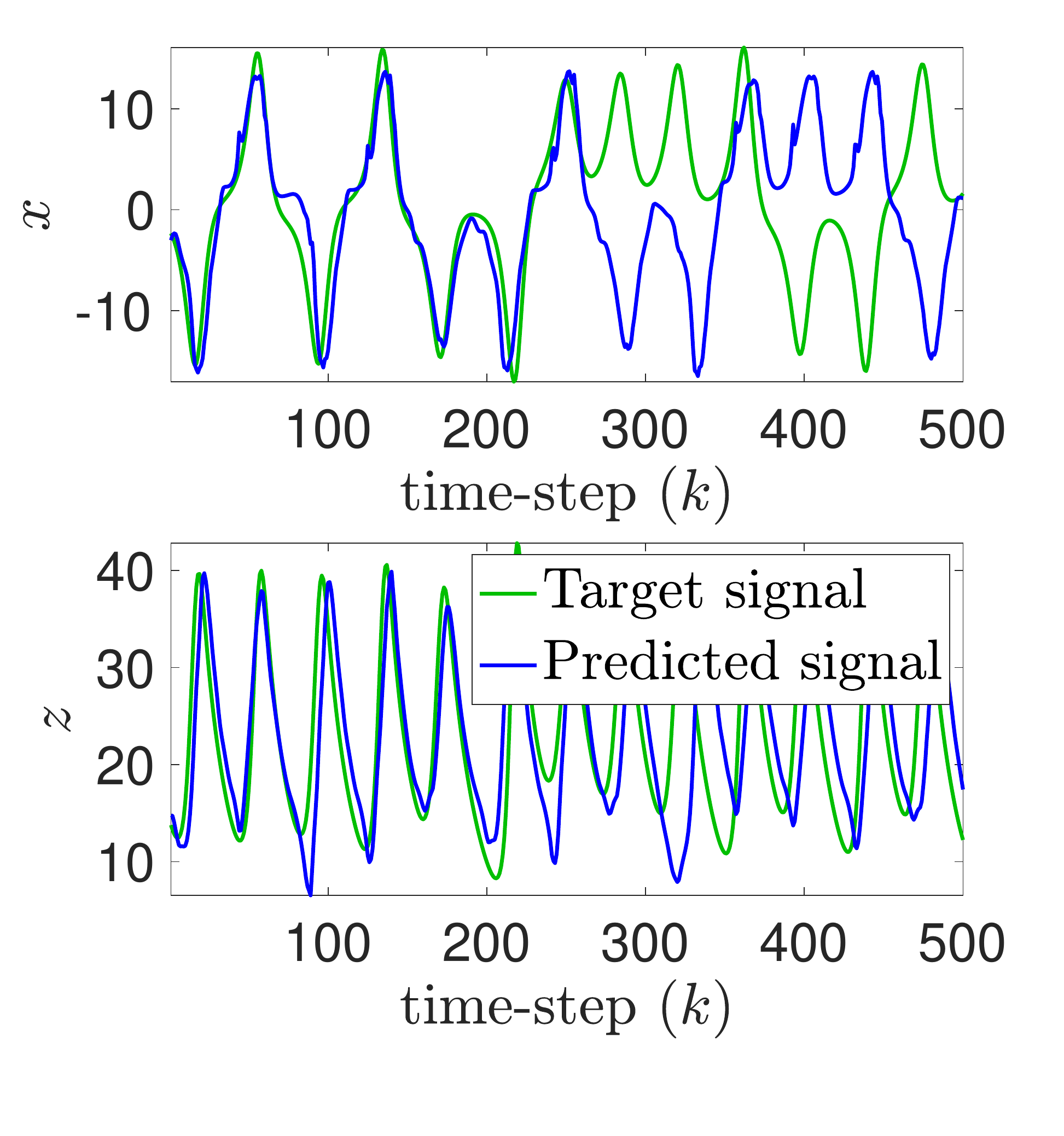}}
\subfloat[]{\includegraphics[trim=0cm 0cm 0cm 0cm, clip=true, width=0.25\textwidth]{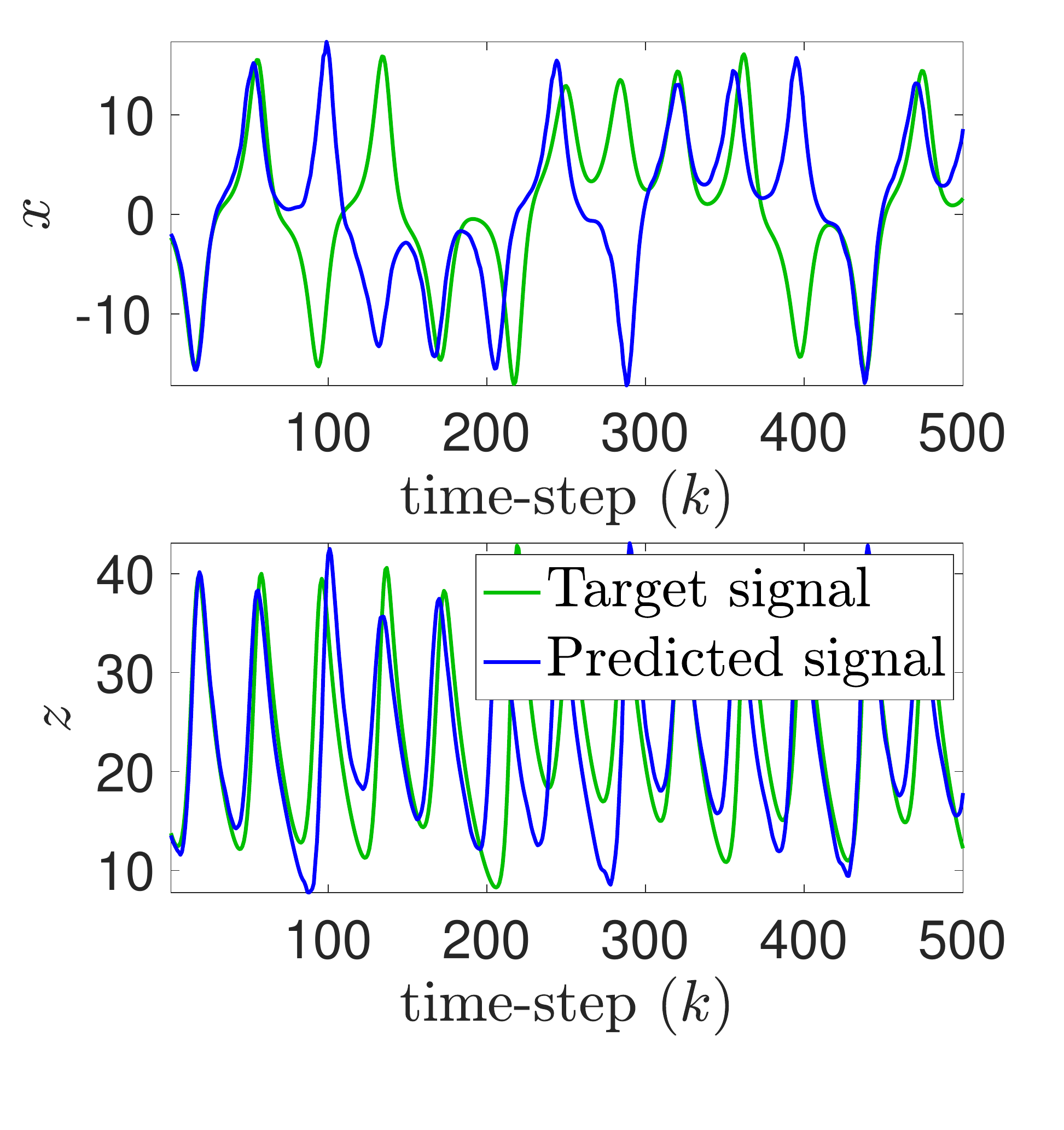}}
\caption{prediction of the noisy time-series $x(t_k)$ and $z(t_k)$ from Lorenz system \eqref{Eq: Lorenz} with $\sigma_v^2 = 1.0$: (a) true and predicted signal with KalT-ESN, (b)  true and predicted signal with least square training} \label{Fig: LorenzPrediction}
\end{figure}

\begin{figure}[t]
\centering 
\includegraphics[trim=1cm 0cm 0cm 0cm, clip=true, width=0.5\textwidth]{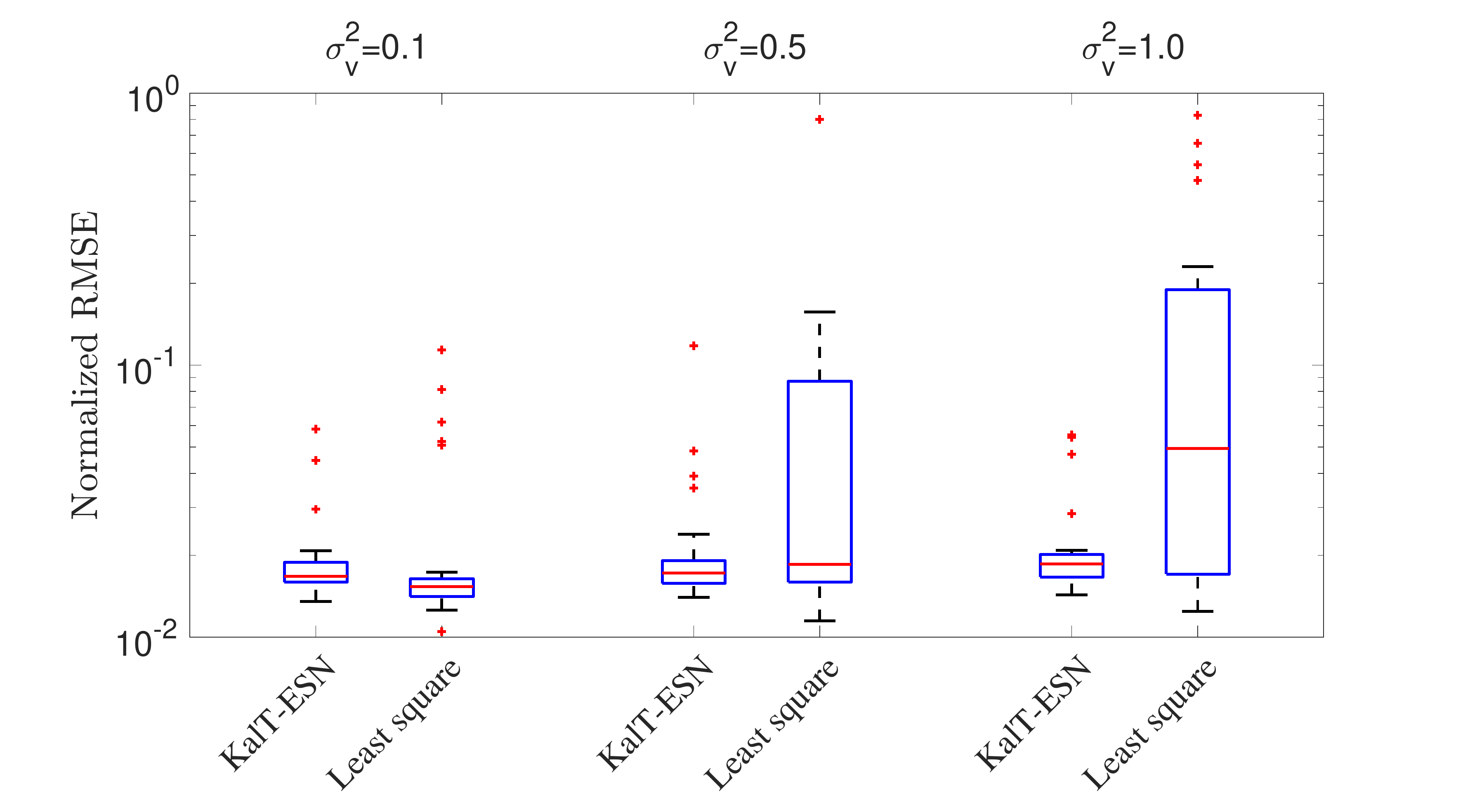}
\caption{Error profile of Lorenz time-series prediction: NRMSE with different measurement noise covariance $\sigma_v^2$} \label{Fig: LorenzError}
\end{figure}

\begin{figure}[t]
\centering 
\subfloat[]{\includegraphics[trim=0cm 0cm 0cm 0cm, clip=true, width=0.25\textwidth]{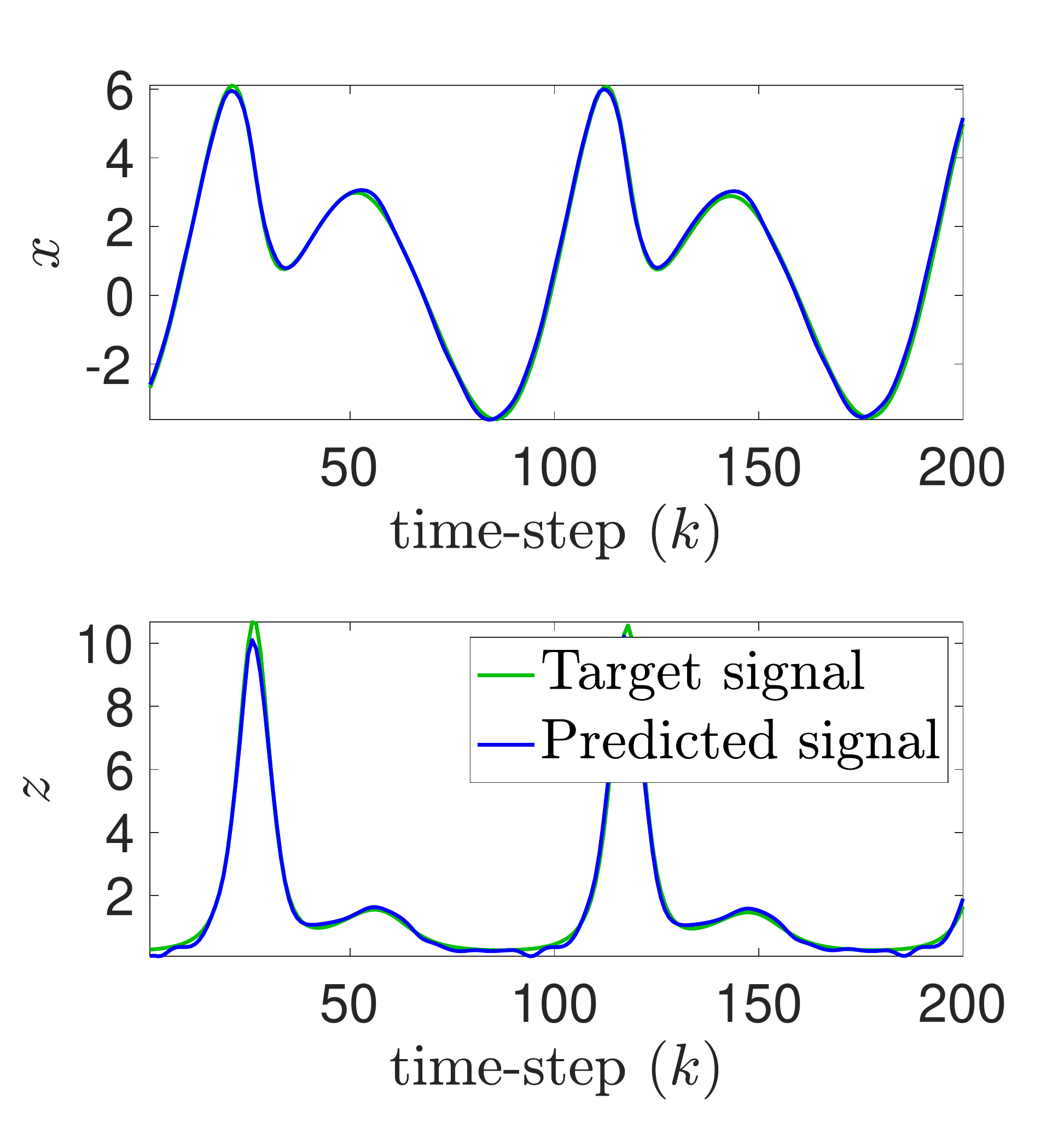}}
\subfloat[]{\includegraphics[trim=0cm 0cm 0cm 0cm, clip=true, width=0.25\textwidth]{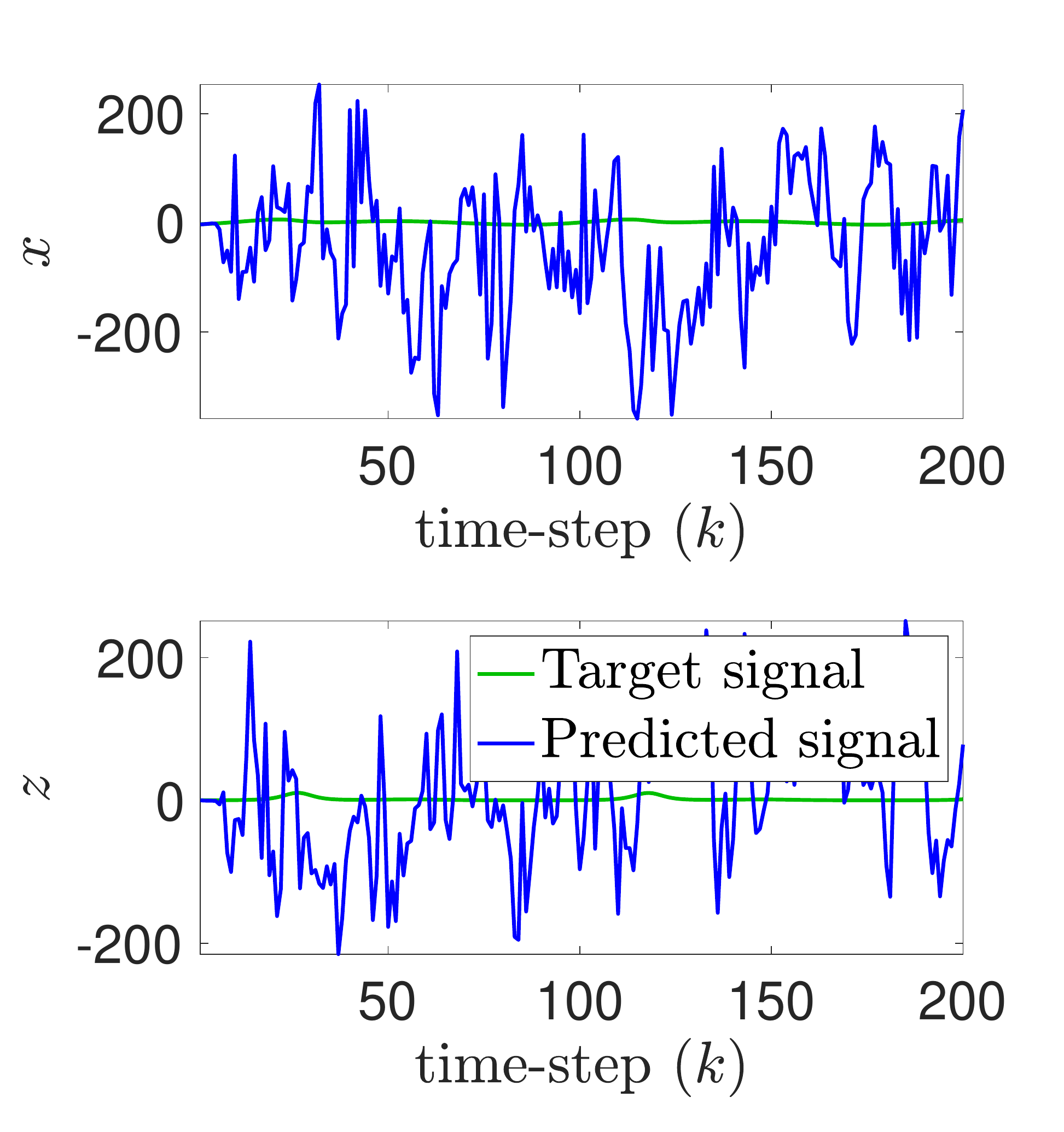}}
\caption{prediction of the noisy time-series $x(t_k)$ and $z(t_k)$ from R\"ossler system \eqref{Eq: Rossler} with $\sigma_v^2 = 0.1$: (a) true and predicted signal with KalT-ESN, (b)  true and predicted signal with least square training} \label{Fig: RosslerPrediction}
\end{figure}

\begin{figure}[t]
\centering 
\includegraphics[trim=1cm 0cm 0cm 0cm, clip=true, width=0.5\textwidth]{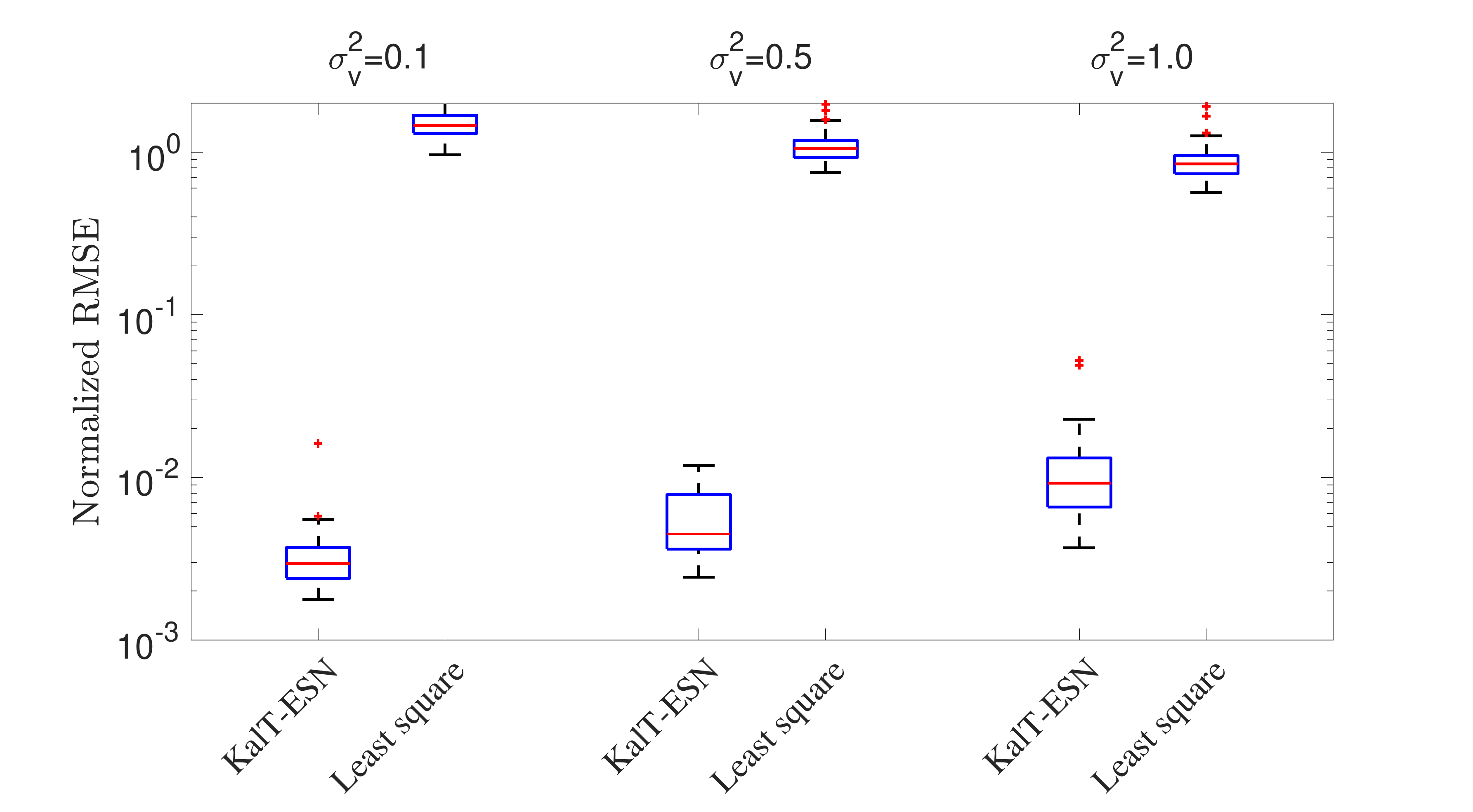}
\caption{Error profile of R\"ossler time-series prediction: NRMSE with different measurement noise covariance $\sigma_v^2$} \label{Fig: RosslerError}
\end{figure}

\begin{figure}[t]
\centering 
\subfloat[]{\includegraphics[trim=1cm 0.5cm 2cm 0.2cm, clip=true, width=0.25\textwidth]{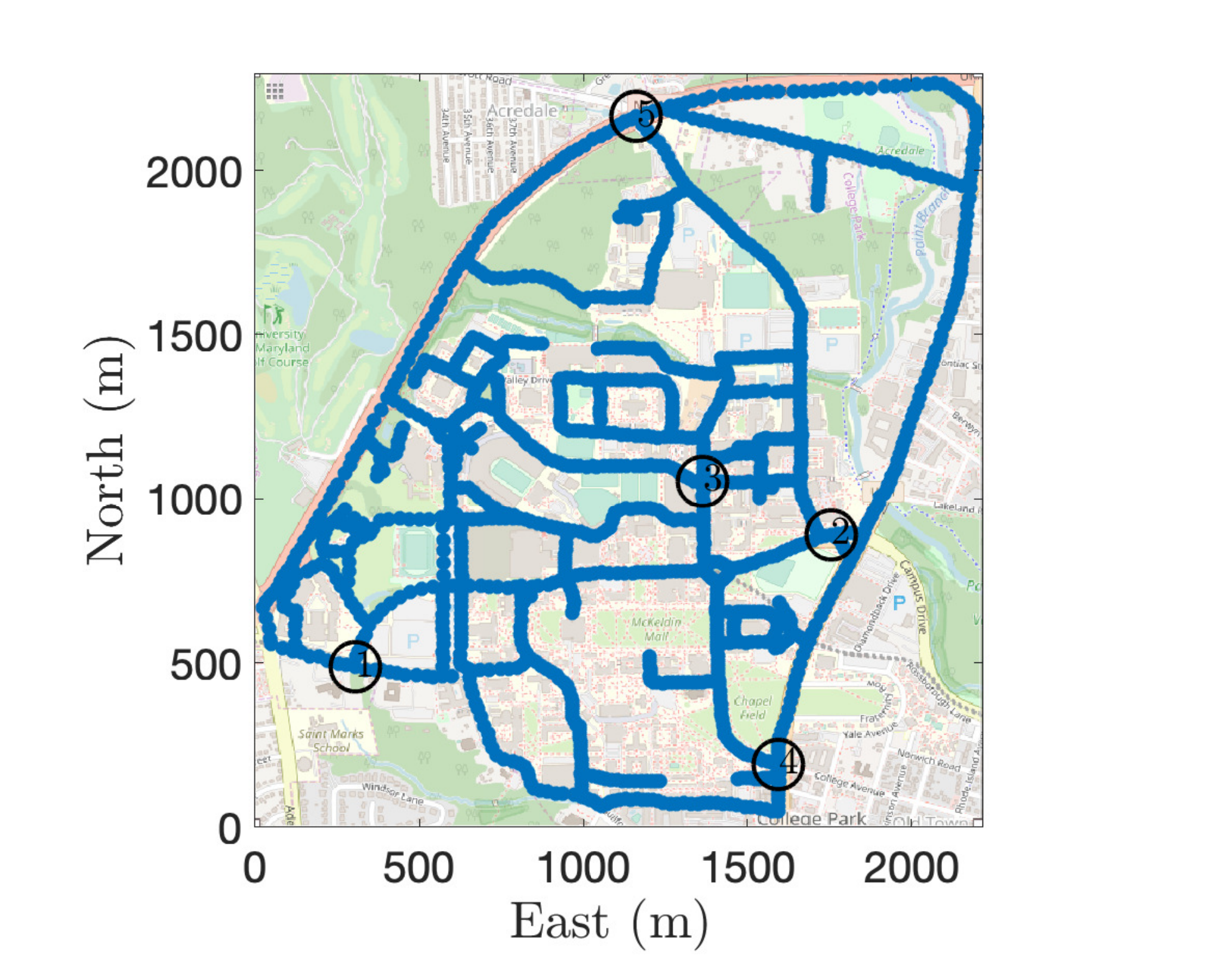}}
\subfloat[]{\includegraphics[trim=2cm 0cm 0cm 0cm, clip=true, width=0.25\textwidth]{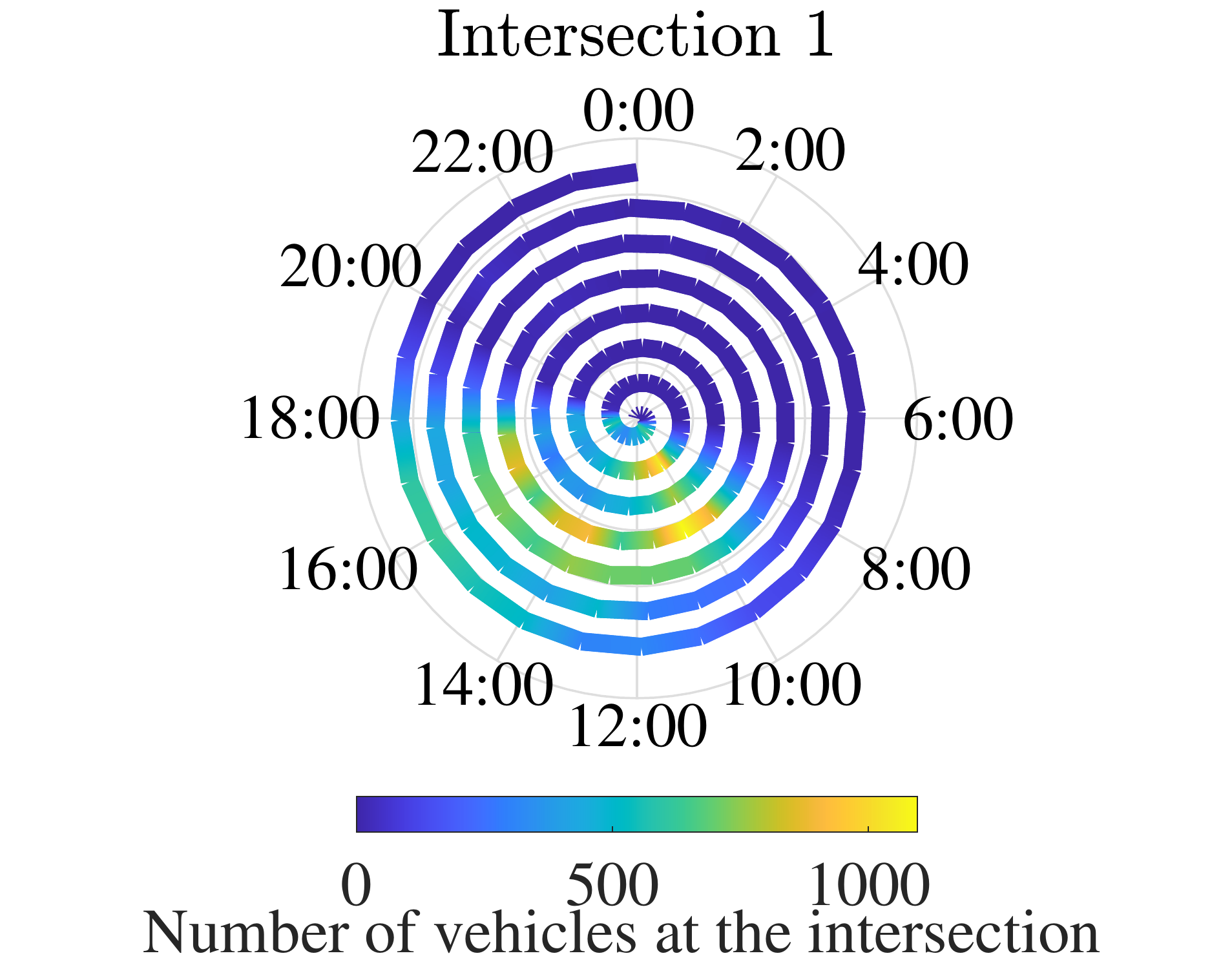}}
\caption{Schematic diagram of traffic data: (a) University of Maryland road network with Numina sensors, (b) Traffic congestion pattern of an intersection over a single week, each revolution denotes a day of the week with times marked as angles; the number of vehicles is denoted by the colormap. The daily pattern of peak congestion between mornings and afternoons is evident.} \label{Fig: TrafficSchematic}
\end{figure}

\begin{figure}[t]
\centering 
\includegraphics[trim=0cm 0cm 0cm 0cm, clip=true, width=0.5\textwidth]{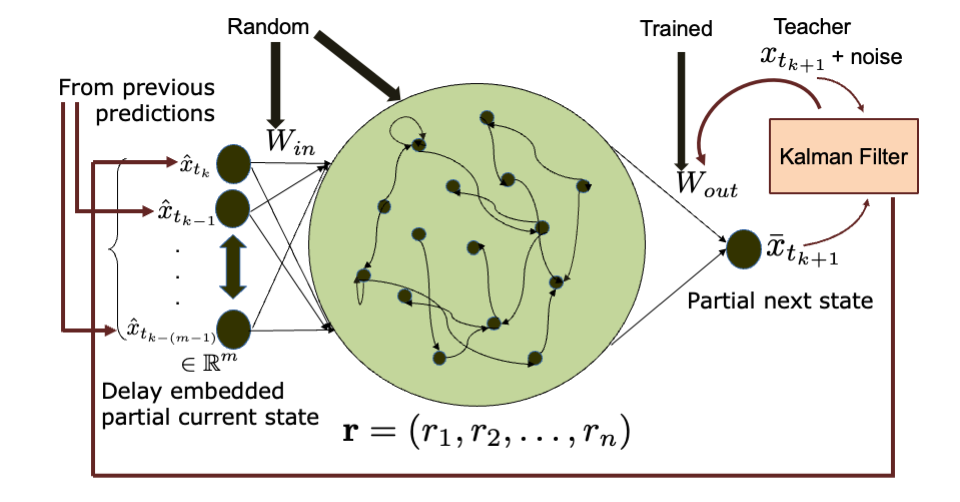}
\caption{$m$-dimensional delay-embedding of the scalar partial observation for KalT-ESN} \label{Fig: DelayESNKalman}
\end{figure}

\begin{figure}[t]
\centering 
\subfloat[]{\includegraphics[trim=0cm 0cm 0cm 0cm, clip=true, width=0.25\textwidth]{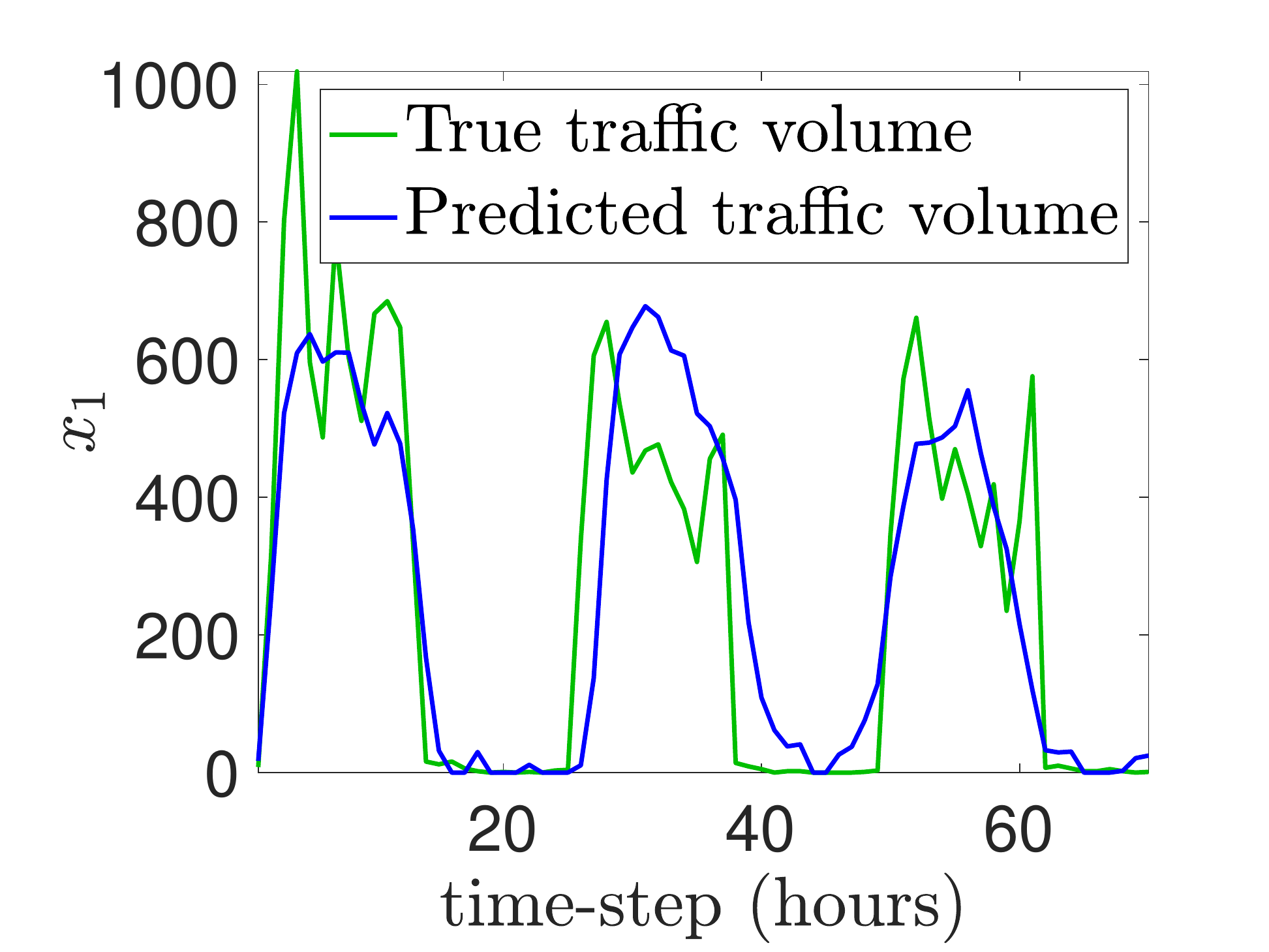}}
\subfloat[]{\includegraphics[trim=0cm 0cm 0cm 0cm, clip=true, width=0.25\textwidth]{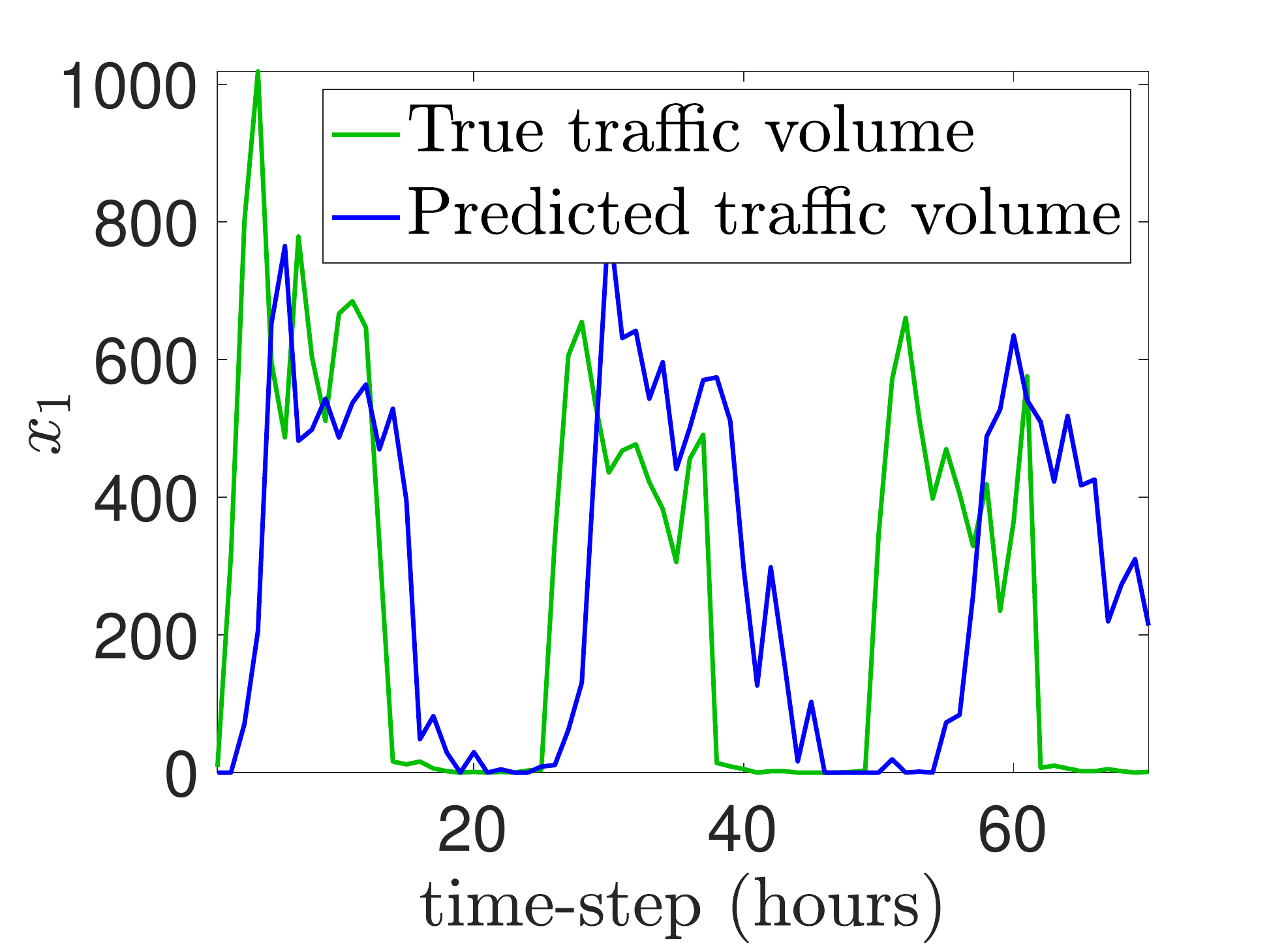}}
\caption{prediction of the noisy time-series of traffic volume recorded in Numina sensor 1 with $\sigma_v^2 = 500$: (a) true and predicted traffic volume with KalT-ESN, (b)  true and predicted traffic volume with least square training} \label{Fig: NuminaPrediction}
\end{figure}

\begin{figure}[t]
\centering 
\subfloat[]{\includegraphics[trim=0cm 0cm 0cm 0cm, clip=true, width=0.25\textwidth]{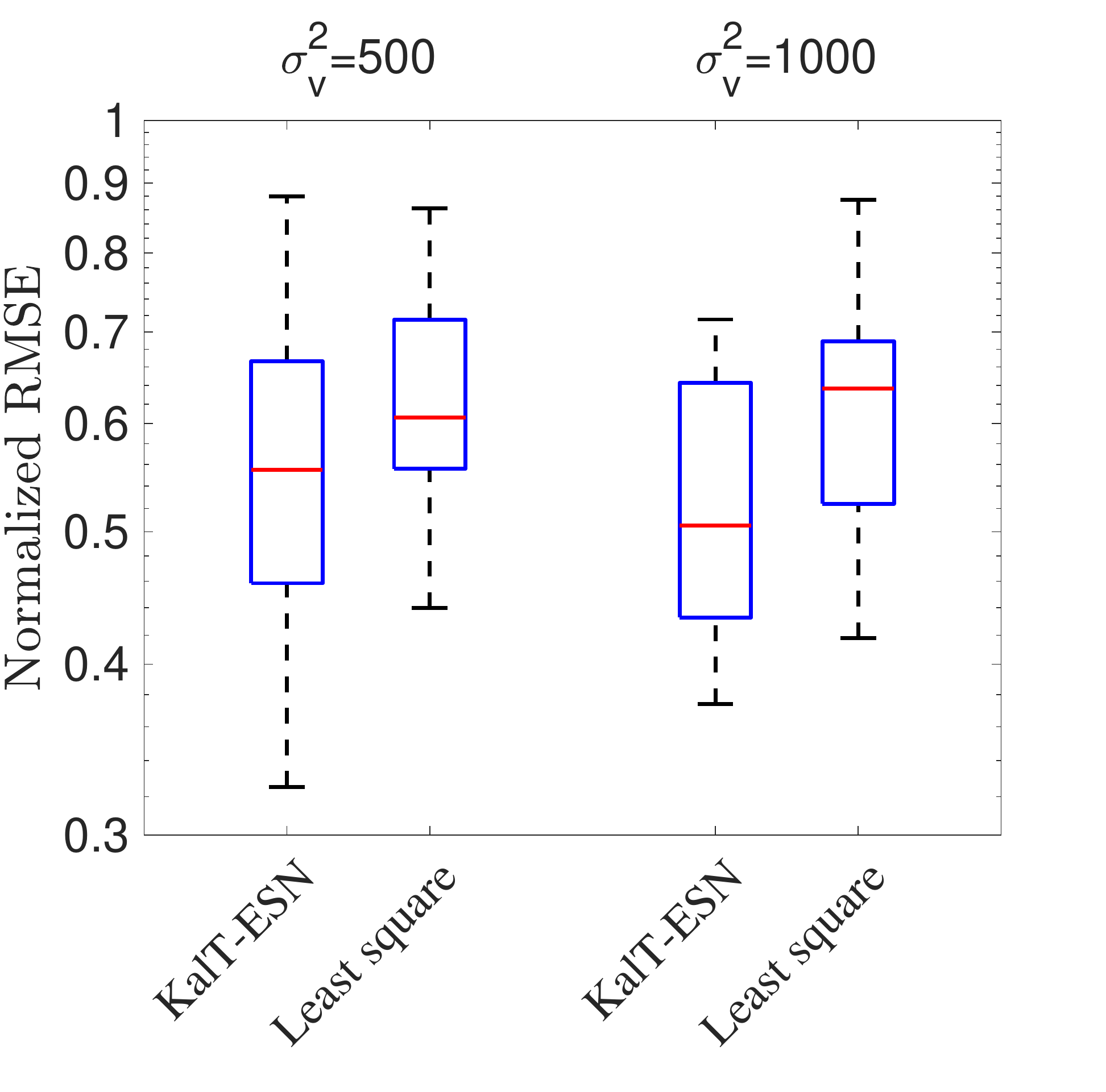}}
\subfloat[]{\includegraphics[trim=0cm 0cm 0cm 0cm, clip=true, width=0.25\textwidth]{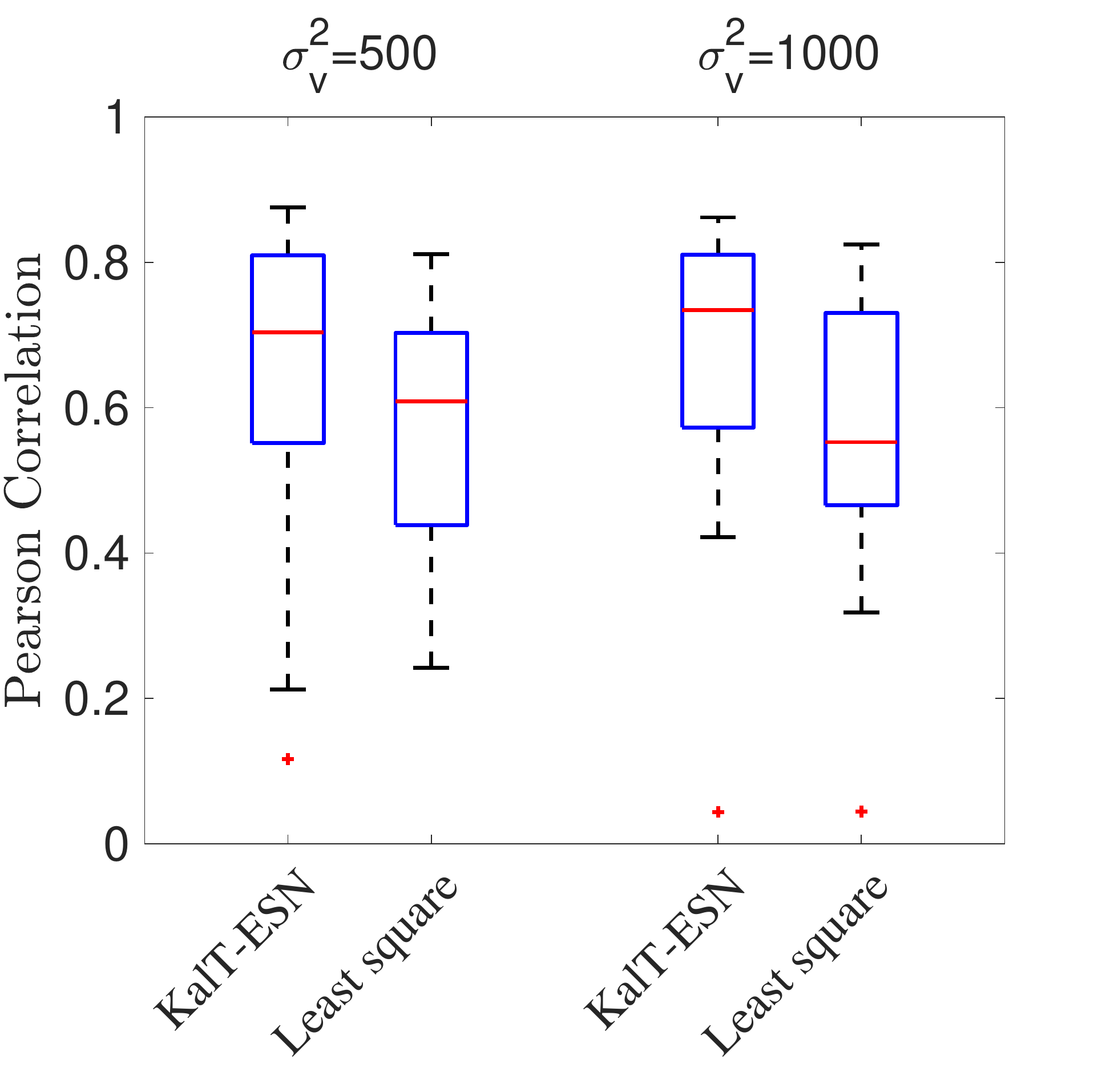}}
\caption{Error and correlation profile of traffic volume prediction: (a) NRMSE  and (b) Pearson correlation with different measurement noise covariance $\sigma_v^2$. The delay embedding dimension is $m=10$.} \label{Fig: NuminaError}
\end{figure}

\subsection{Lorenz System}
The KalT-ESN algorithm is tested on a time-series $[x(t_k)\,\, y(t_k) \,\, z(t_k)]$ generated by the Lorenz system:
\bnl\label{Eq: Lorenz}
\dot{x} &=& \sigma(y-x)\\\nonumber
\dot{y} &=& x(\rho-z) -y \\\nonumber
\dot{z} &=& xy - \beta z,
\enl
where $\sigma=10$, $\rho=28$, and $\beta=8/3$ produces chaotic behavior. The training data is corrupted by a measurement noise $\mf{v}(t_k) \sim \mc{N}(0, \sigma_v^2I_{3\times 3})$. Table \ref{tb:hyparam} lists the hyperparameters used to train the ESN. The prediction via KalT-ESN and least square training is depicted in Fig.~\ref{Fig: LorenzPrediction}. Fig.~\ref{Fig: LorenzError} provides a detailed error profile for measurement noise covariance $\sigma_v^2$. The normalized root mean square error (NRMSE) between the true sequence $\{\mf{x}(t_k): i=1,\ldots, l\}$ and the predicted sequence $\{\hat{\mf{x}}(t_k): i=1,\ldots, l\}$ is given by
\bql
\operatorname{NRMSE}(\mf{x},\hat{\mf{x}}) = \sqrt{\dfrac{\sum\limits_{k=1}^l \norm{\mf{x}(t_k)-\hat{\mf{x}}(t_k)}^2}{\sum\limits_{k=1}^l \norm{\mf{x}(t_k)}^2}},
\eql
where $l$ is the prediction length. The prediction NRMSEs for different measurement noise covariances over 30 independent Monte-Carlo trials are plotted in Fig.~\ref{Fig: LorenzError}. The performance of KalT-ESN algorithm remains consistent with different $\sigma_v$ while the performance of the least square training degrades heavily with higher noise covariance.

\subsection{R\"ossler System}
Next, KalT-ESN is utilized to predict the noisy state measurement generated by the R\"{o}ssler system described in \cite{Rossler1976}:
\bnl\label{Eq: Rossler}
\dot{x} &=& -y-x\\\nonumber
\dot{y} &=& x + ay \\\nonumber
\dot{z} &=& b + z(x-c),
\enl
with $a=0.5$, $b=2$, and $c=4$ to produce chaotic behavior. Similar to the Lorenz system example, the training data is corrupted by a measurement noise $\mf{v}(t_k) \sim \mc{N}(0, \sigma_v^2I_{3\times 3})$. Table \ref{tb:hyparam} lists the hyperparameters used to train the ESN. The prediction via KalT-ESN and least square training is depicted in Fig.~\ref{Fig: RosslerPrediction}. Fig.~\ref{Fig: RosslerError}(b) plots the detailed error profile for measurement noise covariance $\sigma_v^2$. The results are generated by 30 independent Monte-Carlo trials for training and testing the ESNs.

\subsection{Prediction of Traffic Volume on an Intersection of a Road Network}
KalT-ESN is now applied to a dataset of traffic volumes obtained from Numina \cite{Numina} sensors at five different intersections on the University of Maryland campus. Fig.~\ref{Fig: TrafficSchematic}(a) represents the road network marked with sensor locations. Each sensor counts the number of pedestrians, bicycles, and vehicles at the respective intersections and store them in a server. The time series data of hourly vehicle traffic volume for two months is used. Fig.~\ref{Fig: TrafficSchematic}(b) represents the hourly vehicle traffic volume over a week with a clear daily pattern.

The traffic dynamics is an infinite-dimensional spatio-temporal dynamical system evolving over a road network, and hence, the traffic volumes recorded from each sensor provides a partial measurement. An ESN usually requires full state measurements in the training phase \cite{Goswami2021}, \cite{Lu2017}. To mitigate this problem, a delay-embedding in the input layer \cite{Goswami2023} is used for training. The delay-embedding for KalT-ESN is demonstrated in Fig.~\ref{Fig: DelayESNKalman}. Both least square training and KalT-ESN is applied on this delay-embedded time series. Only a noisy scalar measurement from sensor 1 is used in this paper with embedding dimension $m=10$.

The ESN is trained on 500 hours of traffic volume data and tested for 70 hours, i.e., approximately three days. The training hyperparameters are listed in Table \ref{tb:hyparam}. Fig. \ref{Fig: NuminaPrediction} shows the traffic volume prediction by KalT-ESN and least square training. Fig.~\ref{Fig: NuminaError} shows the NRMSE and Pearson correlation coefficient between predicted and true traffic volumes with sensor data from intersection 1 corrupted with noise. The results are similar for the other four intersections and not included here. The Pearson correlation coefficient between true and predicted sequences ($\{x(i): i=1,\ldots, l\}$ and $\{\hat{x}(i): i=1,\ldots, l\}$ respectively) measures their normalized linear correlation. It is given by
\bql
r(x,\hat{x}) = \frac{\sum\limits_{k}\left(x(t_k)-\bar{x}\right)^T\left(\hat{x}(t_k)-\bar{\hat{x}}\right)}{\sqrt{\sum\limits_{i}\norm{x(t_k)-\bar{x}}^{2}} \sqrt{\sum\limits_{k}\norm{\hat{x}(t_k)-\bar{\hat{x}}}^{2}}},
\eql
where $\bar{x}$ and $\bar{\hat{x}}$ denotes the time-average values of $x(t_k)$ and $\hat{x}(t_k)$. KalT-ESN yields improved NRMSE and higher Pearson correlation coefficient with an embedding dimension of $m=10$ only.

\begin{remark}
The delay embedded KalT-ESN provides a practical way to train an ESN from both noisy and partial state measurement, and hence, presents a natural continuation of the delay-embedded ESN presented in \cite{Goswami2023}.
\end{remark}

\section{Conclusion}
This paper proposes a sequential training algorithm for an echo-state network (ESN) that combines the power of universal prediction by an ESN with data-assimilation by ensemble Kalman filter (EnKF). The algorithm, called Kalman training of the echo-state network (KalT-ESN), recursively updates the output weights of an ESN using an EnKF from the noisy training data. The proposed training algorithm demonstrates improved performance in presence of additive noise in the training dataset. It is extended for partial noisy measurements in the training phase using a time-delay embedding at the input layer. The method is then applied to a real data set of traffic patterns on the road network of the University of Maryland, College Park campus to predict the traffic volume at various intersections. For ongoing and future work, inference of unobserved states via time-delay embedded ESN with surrogate spatial interpolation model and a data-driven controller design will be investigated. 
\section*{Acknowledgement}
The author thanks Dr. Derek A. Paley and the University of Maryland Department of Transportation for the Numina sensor data. The author also thanks Dr. Artur Wolek for preprocessing the data.
\balance
\bibliographystyle{IEEEtran}
\bibliography{bibl}

\end{document}